\documentclass[a4paper]{article}
\usepackage[a4paper,left=2.9cm,right=2.4cm,top=3.4cm,bottom=3.4cm]{geometry}
\pdfoutput=1
\usepackage{blkarray}
\usepackage{amsmath}

\usepackage{times}
\usepackage{bm}
\usepackage{natbib}
\usepackage{graphicx,subfigure}
\usepackage{mathrsfs}
\usepackage{mathtools}

\usepackage{amssymb}
\usepackage{amsfonts}
\usepackage{tikz}
\usetikzlibrary{positioning}
\usepackage{dsfont}

\usepackage[plain,noend]{algorithm2e}

\makeatletter
\renewcommand{\algocf@captiontext}[2]{#1\algocf@typo. \AlCapFnt{}#2} 
\def\@algocf@capt@plain{top}
\renewcommand{\algocf@makecaption}[2]{%
  \addtolength{\hsize}{\algomargin}%
  \sbox\@tempboxa{\algocf@captiontext{#1}{#2}}%
  \ifdim\wd\@tempboxa >\hsize
    \hskip .5\algomargin%
    \parbox[t]{\hsize}{\algocf@captiontext{#1}{#2}}
  \else%
    \global\@minipagefalse%
    \hbox to\hsize{\box\@tempboxa}
  \fi%
  \addtolength{\hsize}{-\algomargin}%
}
\makeatother

\newcommand{\QED}{\ifhmode\unskip\nobreak\fi\quad {\rm Q.E.D.}} 

\newcommand{\R}{\mathbb{R}}
\newcommand{\cS}{\mathcal{S}}

\newcommand{\cW}{\mathcal{W}}

\newcommand{\braces}[2]{\genfrac{\{}{\}}{0pt}{1}{#1}{#2}}

\RequirePackage[amsmath,amsthm,thmmarks]{ntheorem}
\theoremheaderfont{\hskip\parindent\normalfont\scshape}
\theorembodyfont{\itshape}
\theoremseparator{.}

\theoremnumbering{arabic}
\newtheorem{theorem}{Theorem}
\theoremnumbering{arabic}

\theoremnumbering{arabic}
\newtheorem{corollary}{Corollary}
\theoremnumbering{arabic}
\newtheorem{proposition}{Proposition}
\theoremnumbering{arabic}
\newtheorem{definition}{Definition}
\theoremnumbering{arabic}
\theoremheaderfont{\hskip\parindent\normalfont\itshape}
\theorembodyfont{\rm}

\theoremnumbering{arabic}

\theoremnumbering{arabic}
\newtheorem{example}{Example}
\theoremnumbering{arabic}

\theoremnumbering{arabic}

\theoremnumbering{arabic}

\theoremnumbering{arabic}

\theoremsymbol{\QEDlogo}
\theoremheaderfont{\hskip\parindent\normalfont\itshape}
\theorembodyfont{\normalfont}
\theoremseparator{.}
\theoremstyle{nonumberplain}
\usepackage{amsmath}

\usepackage{authblk}

\begin{document}




\title{The correlation space of Gaussian latent tree models and model selection without fitting}


\author[1]{N. Shiers\thanks{n.l.shiers@warwick.ac.uk}}
\author[2]{P. Zwiernik\thanks{piotr.zwiernik@gmail.com}}
\author[3]{J. A. D. Aston\thanks{j.aston@statslab.cam.ac.uk}}
\author[1]{J. Q. Smith\thanks{j.q.smith@warwick.ac.uk}}
\affil[1]{Department of Statistics, University of Warwick, Coventry, CV4 7AL, U.K.}
\affil[2]{Department of Economics and Business, Pompeu Fabra University, 08005 Barcelona, Spain}
\affil[3]{Statistical Laboratory, University of Cambridge, Cambridge, CB3 0WB, U.K.}

\renewcommand\Authands{ and }

\maketitle
\begin{abstract}
We provide a complete description of possible covariance matrices consistent with a Gaussian latent tree model for any tree. We then present techniques for utilising these constraints to assess whether observed data is compatible with that Gaussian latent tree model. Our method does not require us first to fit such a tree. We demonstrate the usefulness of the inverse-Wishart distribution for performing preliminary assessments of tree-compatibility using semialgebraic constraints. Using results from \citet{drton_momsWishart} we then provide the appropriate moments required for test statistics for assessing adherence to these equality constraints. These are shown to be effective even for small sample sizes and can be easily adjusted to test either the entire model or only certain macrostructures hypothesized within the tree. We illustrate our exploratory tetrad analysis using a linguistic application and our confirmatory tetrad analysis using a biological application.
\end{abstract}


\section{Introduction}

Modelling with hidden variables is commonly performed within the framework of graphical models \citep{lauritzen:96,koller2009probabilistic}. When the observed variables are the leaves of a tree and the unobserved variables are interior nodes then the model is said to be a latent tree model \citep{choi:2011,wang2008latent}. These models are used across disciplines including sociology, biology, and linguistics \citep{eisenstein2010latent,mourad2013survey,zwiernik_LT}. In the case of continuous data Gaussian latent tree models became a popular choice \citep{lawrence2004gaussian}.

Standard latent tree model selection techniques  often assume a priori that the data generating process is driven by some latent tree model. So in particular the appropriateness of any given tree model is often not assessed in absolute terms but only relative to other tree models. Knowing whether any latent tree model could adequately explain what is observed is pertinent in phylogenetic settings where, for example, the effect of a possible horizontal gene transfer (e.g.\ \citet{hao2008uncovering}) makes any underlying latent tree model hypothesis a dubious one.

By characterising the covariance space related to Gaussian latent tree models, we can better assess the suitability of trees or the fit of a particular tree for a data set. In this paper we present the complete description of this model class by relating this to the space of phylogenetic oranges \citep{toriccubes,gilloranges,kim2000slicing,moulton2004ppo}. Such a complete description had been known for a simple tree with only four leaves (see \citet[Theorem~2]{PearlXu87}) or for a star tree (see \citet{bekker1987rank}). For a general tree, only the defining equations have been derived; see \citet[Corollary~6.5]{sullivant2008agg}.

Our method will use the description of Gaussian latent tree models in two scenarios. In the first setting we are interested in whether any latent tree model is a possible explanation for a given data set. In the second situation we fix a latent tree model. In both situations the alternative hypothesis is given by the saturated model. We illustrate these methods in Section~\ref{sec:cta} where we test a previously hypothesized phylogenetic tree applied to certain yeast species. In both examples it is contentious whether the class of phylogenetic trees is appropriate. This is the question that we are able to address directly in our analyses and something that can be done without first fitting the model.

Let $Z=(Z_u)_{u\in U}$ be a random vector whose components are indexed by the vertices of an undirected tree $T=(U,E)$ with edge set $E\subset U\times U$.  The tree $T$ induces a Gaussian tree model $N(T)$ for $Z$, which is a Gaussian graphical model on $T$ \citep[Section~5.2]{lauritzen:96}.  For any two nodes $u,v\in U$, let ${\rm ph}(uv)$ denote the set of edges on the unique path between $u$ and $v$ in this tree.  Then the model $N(T)$ is the collection of all multivariate normal distributions on $\mathbb{R}^{\vert U\vert}$ for which $Z_u$ and $Z_v$ are conditionally independent given a subvector $Z_C$ whenever the set $C\subset U\setminus\{u,v\}$ contains a node on ${\rm ph}(uv)$.  Note that for three nodes $u,v,w\in U$ the
conditional independence of $Z_v$ and $Z_w$ given $Z_u$ is equivalent
to $\rho_{vw}=\rho_{uv}\rho_{uw}$. It follows that a normal distribution with correlation matrix $R=(\rho_{uv})$ belongs to $N(T)$ if and only if $\rho_{uv}=\prod_{e\in {\rm ph}(uv)} \rho_{e}$ for all $u,v\in U$, where $\rho_{e}=\rho_{uv}$ when $e$ is the edge $(u,v)$.

In this paper we study Gaussian latent tree models where only the observed random variables correspond to the tree's leaves. We henceforth denote the set of leaves of this tree by $V$. A typical such evolutionary tree, one of Romance languages, is displayed below in Fig.~\ref{fig:quintetlang} where the observable, extant languages are represented as its leaves.

 \begin{figure}[ht!]
\centering
\tikzstyle{vertex}=[circle,fill=black,minimum size=5pt,inner sep=0pt]
\tikzstyle{hidden}=[circle,draw,minimum size=5pt,inner sep=0pt]
\begin{tikzpicture}[scale=.9]
  \node[hidden] (h2) []{};  
  \node[hidden] (h1) at (-1.27,0) []{};
  \node[vertex] (x5) at (0,1.27) [label=above:Iberian Spanish]{};
  \node[hidden] (h3) at (1.27,0) []{};
  \node[vertex] (x1) at (-2,1.27) [label=above left:Portuguese]{};
  \node[vertex] (x2) at (-2,-1.27) [label=above left:French]{};
  \node[vertex] (x3) at (2,1.27) [label=above right:American Spanish]{};
  \node[vertex] (x4) at (2,-1.27) [label=above right:Italian]{};
  \path (h1) edge (h2)
		(h2) edge (h3)
		(h1) edge (x1)
 		(h1) edge (x2)
  		(h3) edge (x3)
 		(h3) edge (x4)
  		(h2) edge (x5);
\end{tikzpicture}
\caption{Quintet tree $T_{5}$ relating five Romance languages.} \label{fig:quintetlang}
\end{figure}
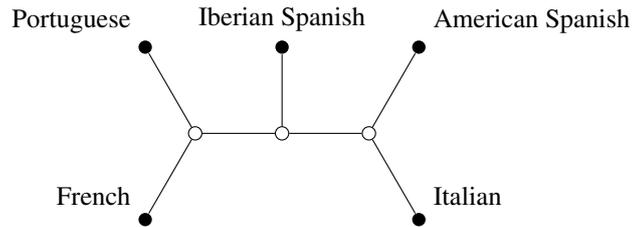

\begin{definition}
The Gaussian latent tree model $M(T)$ for the subvector $X=(Z_v)_{v\in V}$ is the set of all $V$-marginal distributions of the distributions in $N(T)$, where the $V$-marginal distributions are those associated with leaf variables.\end{definition}

The parameterization of $M(T)$ is induced from the parameterization of $N(T)$ and given by
\begin{equation}
  \label{eq:rhoijpath}
  \rho_{ij}\quad=\quad\prod_{e\in {\rm ph}(ij)} \rho_{e}
\end{equation}
$\mbox{for all }i,j\in V$. As the variances $\sigma_{uu}$ for $u\in U\setminus V$ never appear in this parameterization, without loss of generality, we can assume they are equal to $1$.

\section{Semialgebraic description of the latent tree model} \label{sec:treemet}

\subsection{Tree metrics and phylogenetic oranges}

Let $T=(U,E)$ be a tree with leaf set $V\subseteq U$.  Associate to each edge a non-negative number $d_{e}$, which we interpret as the length of this edge. Then for any two leaves $i,j\in V$ we can compute the distance between them as $d_{ij}=\sum_{e\in {\rm ph}(ij)}d_{e}$. It is easy to check that a collection of such distances for all pairs $u,v\in V$ forms a metric. The set of all metrics that arise in this way for all $T$ with leaves labelled by $V$ is called the space of tree metrics. We recall the following result. 
 
 \begin{theorem}[\citet{buneman1974nmp}]\label{th:buneman}A collection of positive numbers $d_{ij}$ for $i,j\in V$ forms a tree metric if and only if for all (not necessarily distinct) $i,j,k,l\in V$ we have
$$
\max(d_{ik}+d_{jl},d_{il}+d_{jk})\geq d_{ij}+d_{kl}.
$$
Equivalently, for any three sums $d_{ik}+d_{jl}$, $d_{il}+d_{jk}$, $d_{ij}+d_{kl}$ two are equal and not less than the third. Moreover, if the above inequalities hold, then generically $T$ is uniquely identified.
 \end{theorem}
In the above theorem the term generically means that the statement holds outside a set of measure zero corresponding to the vanishing of some edge lengths $d_{e}$. We note that a more precise statement is also possible if we allow semi-labelled trees, see \citet[Section~7]{semple2003pol}. A careful analysis shows that this generic tree is always a binary tree, i.e.\ a tree with all its inner nodes of degree three. The usual triangle inequality follows from setting $i,j,k$ distinct and $k=l$ in Theorem~\ref{th:buneman}, which in turn implies that every tree metric is a metric on $V$.

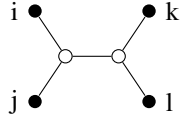
\begin{figure}[ht!]
\centering
\tikzstyle{vertex}=[circle,fill=black,minimum size=5pt,inner sep=0pt]
\tikzstyle{hidden}=[circle,draw,minimum size=5pt,inner sep=0pt]
\begin{tikzpicture}[scale=.5]
  \node[hidden] (h1) at (-1.2,0) []{};
  \node[hidden] (h3) at (0.2,0) []{};
  \node[vertex] (x1) at (-2,1.2) [label=left:i]{};
  \node[vertex] (x2) at (-2,-1.2) [label=left:j]{};
  \node[vertex] (x3) at (1,1.2) [label=right:k]{};
  \node[vertex] (x4) at (1,-1.2) [label=right:l]{};
  \path (h1) edge (h3)
		(h1) edge (x1)
 		(h1) edge (x2)
  		(h3) edge (x3)
 		(h3) edge (x4);
\end{tikzpicture}
\caption{A quartet tree $ij|kl$} \label{fig:quartet}
\end{figure}

\begin{corollary}\label{cor:buneman}The space of tree metrics on a fixed tree $T$ is given as a set of all metrics on $V$ satisfying: for any four distinct leaves $i,j,k,l$ such that ${\rm ph}(i,j)\cap {\rm ph}(k,l)=\emptyset$, we have
$$
d_{ik}+d_{jl}\quad =\quad d_{il}+d_{jk}\quad \geq\quad  d_{ij}+d_{kl}.
$$
\end{corollary}
We emphasize that ${\rm ph}(i,j)$ is the set of edges and hence, for example, for a star tree any four leaves $i,j,k,l$ that satisfy ${\rm ph}(i,j)\cap {\rm ph}(k,l)=\emptyset$. The condition ${\rm ph}(i,j)\cap {\rm ph}(k,l)=\emptyset$ implies that the induced subtree over $i,j,k,l$, that is, the smallest connected subgraph of $T$ containing $i,j,k,l$, looks like a quartet tree in Figure~\ref{fig:quartet}. This also explains the conditions of Corollary \ref{cor:buneman}.

Another closely related space defined over a tree is the space of phylogenetic oranges; see e.g. \citet{kim2000slicing,moulton2004ppo}. For a fixed tree $T$ this is given by the same parameterization (\ref{eq:rhoijpath}) as the Gaussian latent tree model but where in addition the edge correlations $\rho_{e}$ are non-negative. The set of all points in $\R^{m(m-1)/2}$ that arise in this way is denoted by ${\rm PO}(T)$ and it forms a toric cube as defined in \citet{toriccubes}. The union of all ${\rm PO}(T)$ is denoted by ${\rm PO}(V)$. 

Denote by ${\rm PO}_{+}(T)$ and  ${\rm PO}_{+}(V)$ the subsets of ${\rm PO}(T)$ and  ${\rm PO}(V)$ respectively, where all coordinates are assumed to be strictly positive. This implies in particular that the corresponding edge correlations $\rho_{e}$ must be strictly positive. The space of tree metrics on a fixed tree $T$ is isomorphic to ${\rm PO}_{+}(T)$, with the isomorphism given by $d_{ij}=-\log(\rho_{ij})$. 
\begin{theorem}\label{thm:ineq}Let $R=(\rho_{ij})_{i,j\in V}$ and suppose that $\rho_{ij}\geq 0$ for all $i,j\in V$. The following two statements hold:

(1) $R\in {\rm PO}(V)$ if and only if for every four not necessarily distinct elements $i,j,k,l$ in $V$ at least two out of three products
$\rho_{ik}\rho_{jl}$, $\rho_{il}\rho_{jk}$, $\rho_{ij}\rho_{kl}$ are equal and less than or equal to the third. Moreover, if this holds then $T$ with the property $R\in {\rm PO}(T)$ is generically identified uniquely.

(2) For a fixed $T$, the space ${\rm PO}(T)$ has dimension $|E|$. This is described by the following set of constraints. For any four distinct elements $i,j,k,l$ of $V$ such that \mbox{${\rm ph}(i,j)\cap {\rm ph}(k,l)=\emptyset$}, we have that 
\begin{equation}\label{eq:tetrad1}
\rho_{ik}\rho_{jl}\quad =\quad \rho_{il}\rho_{jk}\quad \leq\quad  \rho_{ij}\rho_{kl}.
\end{equation}
Moreover, for any three distinct leaves $i,j,k$ we have the triangle inequality $\rho_{ij}\rho_{ik}\leq\rho_{jk}$.
\end{theorem}



\subsection{Latent tree models and phylogenetic oranges}\label{sec:gltm}

We are now ready to derive the semialgebraic description of the model $M(T)$. Let $\cS_+(V)$ denote the space of all symmetric positive definite $|V|\times |V|$-matrices.

\begin{theorem}\label{th:ineqpos}
Let $T$ be a tree and let  $R=[\rho_{ij}]\in \cS_{+}(V)$ be a correlation matrix. Then $R\in M(T)$ if and only if $R'=[|\rho_{ij}|]\in {\rm PO}(T)$ and $\rho_{ij}\rho_{ik}\rho_{jk}\geq 0$ for any three distinct  $i,j,k\in V$.  
\end{theorem}
The proof is given in the appendix.

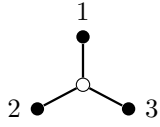
\begin{figure}[ht!]
\centering
\tikzstyle{vertex}=[circle,fill=black,minimum size=5pt,inner sep=0pt]
\tikzstyle{hidden}=[circle,draw,minimum size=5pt,inner sep=0pt]
\begin{tikzpicture}[scale=.5]
  \node[hidden] (v) {};  
  \node[vertex] (2) at (-1.20,-0.64) [label=left:$2$]{};
  \node[vertex] (1) at (0,1.27) [label=above:$1$]{};
  \node[vertex] (3) at (1.20,-0.64) [label=right:$3$]{};
  \draw[line width=.3mm] (v) to (2);
  \draw[line width=.3mm] (v) to (3);
  \draw[line width=.3mm] (v) to (1);  
\end{tikzpicture}
\caption{Tripod tree.} \label{fig:tripod}
\end{figure}

\begin{example}\label{ex:tripod}Let $T$ be the tripod tree in Fig.\ref{fig:tripod}. The space of correlation matrices in  $M(T)$ is described by $$\rho_{12}\rho_{13}\rho_{23}\geq 0, \quad|\rho_{12}\rho_{13}|\leq |\rho_{23}|, \quad|\rho_{12}\rho_{23}|\leq |\rho_{13}|,\quad |\rho_{13}\rho_{23}|\leq |\rho_{12}|.$$ 
If $\rho_{12},\rho_{13},\rho_{23}\geq 0$ then by Theorem~\ref{thm:ineq}(2) the space described by the above inequalities corresponds to ${\rm PO}(T)$. There are three other sign patterns for $\rho_{12}, \rho_{13}, \rho_{23}$ that ensure that $\rho_{12}\rho_{13}\rho_{23}\geq 0$. For every such pattern we obtain a copy of ${\rm PO}(T)$. Quite remarkably, the space of the correlation matrices in $M(T)$ looks exactly like the three-dimensional slice of the corresponding binary latent class model; see \citet[Figure 1]{ARSZ2013}. It is interesting to note that such constraints cannot, in general, be neglected. For example, simple calculations show that the ratio of the volume of $M(T)$ to the volume of all $3\times 3$ correlation matrices is only $\frac{2}{\pi^{2}}\approx 0.2$. 
\end{example}

Based on Theorem \ref{thm:ineq}(2) and Theorem \ref{th:ineqpos} we formulate the following result.
\begin{proposition}\label{prop:fixedpot}
If $T$ is a fixed tree then the space $M(T)$ has dimension $|V|+|E|$. Let $\Sigma$ be a covariance matrix with no zeros. Then $\Sigma\in M(T)$ if and only if for any three distinct leaves $i,j,k$
\begin{equation}\label{eqn:prod3}
(\sigma_{kk}\sigma_{ij}-\sigma_{ik}\sigma_{jk})(\sigma_{jj}\sigma_{ik}-\sigma_{ij}\sigma_{jk})(\sigma_{ii}\sigma_{jk}-\sigma_{ij}\sigma_{ik})\geq 0,
\end{equation}
and for any four distinct elements $i,j,k,l$ of $V$ such that ${\rm ph}(i,j)\cap {\rm ph}(k,l)=\emptyset$ \begin{equation}\label{eqn:tetrad}
\frac{\sigma_{ik}\sigma_{jl}}{\sigma_{ij}\sigma_{kl}}\quad=\quad\frac{\sigma_{il}\sigma_{jk}}{\sigma_{ij}\sigma_{kl}}\quad\leq\quad 1.
\end{equation}
\end{proposition}
This full algebraic and semialgebraic description can be viewed as a generalization from star trees to general trees of the main results in \citet{bekker1987rank,PearlXu87}. An analogous description of the second order moments for binary latent tree models was given in \cite{pwz-2009-semialgebraictrees}. The similarity of both descriptions comes from the fact that the parameterization of correlations in the binary latent tree model is precisely (\ref{eq:rhoijpath}); see \citet[Lemma 4.1]{pwz-2009-semialgebraictrees}. 

\subsection{Necessary constraints for non-Gaussian tree models}

Fix an inner node $r$ in a tree $T$ and direct all edges away from $r$ to obtain a rooted tree $T^r$. Set 
\begin{equation}\label{eq:linear}
Z_r= \epsilon_r \quad \mbox{ and } \qquad Z_v=\lambda_v Z_u+\epsilon_v\quad\mbox{for} \quad u\to v\mbox{ in }T^r,	
\end{equation}
where $\lambda_v\in \R$ and $\epsilon_v$ for $v\in U$ are independent random variables with mean zero. It is known that the vector $Z$ follows the Gaussian tree model on $T$ if and only if all $\epsilon_v$ are Gaussian. Its subvector $X$ corresponding to the leaf nodes will then follow the latent tree model on $T$. 

It is natural to ask what happens if $\epsilon_v$ are not jointly Gaussian. In a nonparametric setting we could instead assume that $\epsilon_v$ have distributions in the family of all univariate distributions with mean zero and finite variance. In this case it is easily shown that all second order moments of $Z$ exist, $\lambda_v={\rm cov}(Z_u,Z_v)/{\rm var}(Z_u)$, and the correlations satisfy the parameterization in (\ref{eq:rhoijpath}). In particular, we have the following result that provides a set of necessary constraints on the correlations of a non-Gaussian tree model. 
\begin{theorem}
Suppose that $Z$ is a random vector satisfying the recursive equations in (\ref{eq:linear}) for a rooted tree. If all $\epsilon_v$ have finite variance, then the correlation matrix of $X$ must satisfy the constraints of the Gaussian latent tree model.  	
\end{theorem}
So it appears that our results, whilst focused on Gaussian modes, apply to and could be extended beyond this setting.

\section{Utilising semialgebraic constraints} \label{sec:ineqs}

We now describe how the semialgebraic constraints can be used more formally to give an indication of tree-compatibility. Here the constraints in (\ref{eqn:prod3}), which hold for every tree topology, will be called tree-compatibility constraints. A test based on these constraints can be used as an effective preliminary assessment tool to inform whether it is legitimate to proceed to a more complex tetrad analysis. For a fixed $T$ we can further extend our test by including the inequality constraints in (\ref{eqn:tetrad}). The resulting constraints are called $T$-compatibility constraints. A test of fit based on such constraints is called a tree-compatibility or $T$-compatibility test as appropriate. 

A straightforward but effective assessment of $T$-compatibility constraints can be obtained from the posterior probabilities by applying an inverse-Wishart prior on the sample covariance. More precisely, if $\hat{\Sigma}$ is a sample covariance matrix based on a sample $X$ of size $n$ from $\mathcal N_{m}(0,C)$, then the estimated scatter matrix is calculated as ${S} = n\hat{\Sigma} = XX^{T}$ and it is well known that the scatter matrix is Wishart distributed ${S} \sim \mathcal W_{m}(n,C)$ \citep{wishart1928generalised}. A common prior distribution for unknown covariance $C$ is the inverse-Wishart $\mathcal W_{m}^{-1}(n_{0},C_{0})$, e.g. \citet{gel13bda,car08bmd, roverato2002}. The inverse-Wishart is a conjugate prior and so the posterior density $p(C\mid X)$ is inverse-Wishart  $\mathcal W_{m}^{-1}(n_{0}+n,C_{0}+{S})$.
As in \citet{roverato2002}, for $C_{0}$ the identity matrix $I_{\vert V \vert}$ can be used and by letting $n_{0} = m$ ensure that the prior density is well defined. Then $C\mid X$ can be sampled with each draw being translated to a correlation and then tested against the constraints. After $N$ such draws from the posterior distribution an estimate of the posterior probability that $C$ satisfies the positivity constraint can be obtained. Of course other choices of families of priors could be chosen instead (for example the scaled inverse-Wishart \citep{o2008domain}) or we could use a strategy that models correlation and covariance separately \citep{barnard2000modeling}. However, these alternatives bring additional computational cost and complexity. Alternatively, it may be possible to adapt the work on inequality-constrained hypotheses to this framework using Bayesian methods, see \citet{van2012bayesian, gu2014bayesian, gardner2014bayesian}.

In Example~\ref{ex:tripod}, an estimate of the probability of $C$ satisfying the semialgebraic structure of $M(T)$ can be constructed using indicator functions. For each draw $l$ from the relevant inverse-Wishart posterior distribution for $\hat{\Sigma}$, the following identity is evaluated:
\begin{equation}\label{eq:indicl}
r^{l}_{123}(\hat{\Sigma}) = \mathds{1}{\{(\tilde\sigma_{33}\tilde{\sigma}_{12}-\tilde{\sigma}_{13}\tilde{\sigma}_{23})(\tilde \sigma_{22}\tilde{\sigma}_{13}-\tilde{\sigma}_{12}\tilde{\sigma}_{23})(\tilde\sigma_{11}\tilde{\sigma}_{23}-\tilde{\sigma}_{12}\tilde{\sigma}_{13})\geq 0\}}
\end{equation}
where $\tilde{\sigma}_{ij}, i,j = 1,2,3$ are the covariances corresponding to covariance draw $l$ of the posterior, the index $l$ being dropped to keep the notation clean. The posterior probability of tree-compatibility is thus estimated using:
\begin{equation}\label{eq:indic}
R_{123}(\hat{\Sigma}) = \frac{1}{N}\sum_{l=1}^{N} r^{l}_{123}(\hat{\Sigma})
\end{equation}
For a tree with four variables such that ${\rm ph}(1,2)$ and ${\rm ph}(3,4)$ do not intersect, the final test of the inequality constraints is then:
\begin{equation} \label{eq:tetradineq}
R_{12|34}(\hat{\Sigma}) \;=\;  \frac{1}{N} \sum_{l=1}^{N} 
\mathds{1}{\{\tilde{\sigma}_{14}\tilde{\sigma}_{23}-\tilde{\sigma}_{12}\tilde{\sigma}_{34}\leq 0\}}\mathds{1}{\{\tilde{\sigma}_{13}\tilde{\sigma}_{24}-\tilde{\sigma}_{12}\tilde{\sigma}_{34}\leq 0\}}\prod_{\substack{1\leq i<j\\ <k \leq 4}} r^{l}_{ijk}(\hat{\Sigma})
\end{equation}
These sampling approaches do not extend to the algebraic constraints because the set of draws from the posterior satisfying an equality constraint will have zero probability. Thus an alternative approach is needed that uses sample distributions of the minors of a covariance matrix.

\section{The sample distribution of algebraic constraints}\label{sec:moments}
From the previous section and Theorem \ref{thm:ineq}(2), the signs of tetrad constraints $\sigma_{ik}\sigma_{jl}-\sigma_{il}\sigma_{jk}$ and other quadratic binomials of the form $\sigma_{ii}\sigma_{jk}-\sigma_{ij}\sigma_{ik}$ provide essential information about whether a Gaussian distribution lies in $M(T)$. This type of constraints can be realized as minors of the covariance matrix $\Sigma$, that is
\begin{equation}\label{eq:asMinors}
\det(\Sigma_{ij,kl}),\qquad\det(\Sigma_{ij,ik}),
\end{equation}
where $\Sigma_{ij,kl}$ denotes the $ 2\times 2$ sub-matrix of $\Sigma$ with rows $i$ and $j$ and columns $k$ and $l$.
Let $\braces{m}{2}$ denote the set of all subsets of $\{1,\ldots,m\}$ of cardinality two.  We now propose the following estimator of the value of $\det(C_{I,J})$ for $I,J\in \braces{m}{2}$
\begin{equation}\label{eq:Qij}
Q_{I,J}\quad=\quad \frac{1}{n(n-1)}\det(S_{I,J}).
\end{equation}
We note from \citet[Corollary~4.2]{drton_momsWishart} that $Q_{I,J}$ is an unbiased estimator of $\det (C_{I,J})$.   

In what follows we provide the covariances between different $Q_{I,J}$. 
It is convenient to introduce the following notation. For an $m\times m$ matrix $A$ let $A^{(2)}$ denote the matrix with rows and columns indexed by elements $\braces{m}{2}$ whose $(I,J)$-th element is the corresponding minor $\det(A_{I,J})$. With this notation, the matrix, whose elements are the estimators $Q_{I,J}$, is $S^{(2)}/(n(n-1))$.

Sadly, there is no simple explicit formula for covariances of various $2$-minors. However, these can be computed if the true distribution $C$ is known. From \citet[Proposition~3.3]{drton_momsWishart}
\begin{equation}\label{eq:covwhite}
{\rm cov}(S^{(2)})=\{(C^{1/2})^{(2)}\otimes (C^{1/2})^{(2)}\}\{{\rm cov}(W^{(2)})\}\{(C^{1/2})^{(2)}\otimes (C^{1/2})^{(2)}\},
\end{equation}
where $W$ has standard Wishart distribution $\cW_{m}(n,I)$ and $\otimes$ is the Kronecker product.

In the rest of this section we provide a complete description of the covariance matrix ${\rm cov}(W^{(2)})$. Our discussion follows \citet[Example~4.6]{drton_momsWishart}. This gives the same derivation for the case $m=4$. We show below that the generalization to $m\geq 4$ is straightforward. 

The matrix ${\rm cov}(W^{(2)})$ has many symmetries that we want to exploit. Note that for all $I,J\in \braces{m}{2}$, $\det W_{I,J}=\det W_{J,I}$ and hence
$$
{\rm cov}\{\det (W_{I,J}),\det (W_{K,L})\}={\rm cov}\{\det (W_{J,I}),\det (W_{K,L})\}={\rm cov}\{\det (W_{K,L}),\det (W_{I,J})\}.
$$
We can therefore, without loss, consider only unordered pairs of sets $(I,J)$, where $I=\{i,j\}$ and $J=\{k,l\}$ with $i<j$, $k<l$ and either $i<k$ or $i=k$ and $j\leq l$.

 Let  $A\Delta B=(A\setminus B)\cup (B\setminus A)$ be the symmetric difference of two sets. We split the rows and the columns of ${\rm cov}(W^{(2)})$ into blocks according to the value of $I\Delta J$ and $K\Delta L$. With this convention, by \citet[Corollary~4.2 and Proposition~3.4]{drton_momsWishart}, ${\rm cov}(W^{(2)})$ is a block-diagonal matrix. Therefore, it is enough to describe its diagonal blocks. Since $|I\Delta J|\in \{0,2,4\}$, we have three types of blocks. We first describe the block corresponding to $I\Delta J=K\Delta L=\emptyset$, or equivalently $I=J$, $K=L$. This block forms a ${m\choose 2}\times {m\choose 2}$-matrix that satisfies:
\begin{equation*}
{\rm cov}\{\det (W_{I,I}),\det (W_{K,K})\}=\left\{\begin{array}{ll}
0 &\qquad |I\cap K|=0\\
2n(n-1)^2 &\qquad |I\cap K|=1\\
2n(2n+1)(n-1) &\qquad I=K.
\end{array}\right.
\end{equation*} 
We now have ${m\choose 2}$ blocks corresponding to $I\Delta J=K\Delta L=\{i,j\}$ for $1\leq i<j\leq m$. Every such block is an $(m-2)\times (m-2)$-matrix, where $I=\{i,k\}$, $J=\{j,k\}$, $K=\{i,l\}$, $L=\{j,l\}$ for some $k\leq l\in \{1,\ldots,m\}\setminus \{i,j\}$. All of these matrices have two types of elements. The diagonal entries ($k=l$) are equal to $n(n+2)(n-1)$. The off-diagonal elements ($k<l$), up to a sign, are $n(n-1)^2$. The sign depends on the relative order of $i,j,k,l$. By \citet[Theorem~4.5]{drton_momsWishart}, the sign is positive if $k<i<j<l$. Now a simple sign analysis shows that the sign is negative only if either $i<k<j<l$ or $k<i<l<j$. This yields that:
\begin{equation*}
{\rm cov}\{\det (W_{ik,jk}),\det (W_{il,jl})\}=\left\{\begin{array}{ll}
n(n+2)(n-1)\quad & k=l\\
-n(n-1)^2\quad & i<k<j<l\mbox{ or }k<i<l<j\\
n(n-1)^2\quad & \mbox{otherwise}.
\end{array}\right.
\end{equation*}

Finally, there are ${m\choose 4}$ blocks corresponding to $I\Delta J=\{i,j,k,l\}$, where $1\leq i<j<k<l\leq m$. Each such block is a $3\times 3$ matrix of the form
\[
\begin{blockarray}{ccc}
ij,kl & ik,jl & il,jk  \\
\begin{block}{[ccc]}
  a & b & -b \\
  \cdot & a & b \\
  \cdot & \cdot & a  \\
  \end{block}
\end{blockarray}
 \]where $a=2n(n-1)$ and $b=n(n-1)$.

\section{Quartets and applications of tetrad analyses} \label{sec:quart}

\subsection{The method of quartets}
\label{sec:quarttest}
For any four distinct leaves $i,j,k,l\in V$ we say that $q_{ij,kl} = ij\vert kl$ forms a quartet of $T$ if the paths ${\rm ph}(i,j)$ and ${\rm ph}(k,l)$ are disjoint, c.f. Figure~\ref{fig:quartet}. A binary tree $T$ displays the set of quartets $\mathcal{Q}$ if each quartet $q \in \mathcal{Q}$ is a quartet of $T$. A set of quartets $\mathcal{Q}$ is said to determine $T$ if $T$ displays $\mathcal{Q}$ and $T$ is the unique tree displayed by $\mathcal{Q}$ \citep{semple2003pol}; the set of all quartets displayed by $T$ is denoted by $\mathcal{Q}_{T}$. Quartets can be considered as fundamental components of binary trees; see  \citet{bookPhylogeneticCombinatorics2012} for more details. A set $\mathcal{Q}_{T}$ is said to be minimal if there exists no element $q \in \mathcal{Q}_{T}$ such that $\mathcal{Q}_{T}\setminus\{q\}$ determines $T$. \citet[Theorem~~2.4]{grunewald2008quartet} provides the minimum size of any $\mathcal{Q}_{T}$ (i.e.\ the size of the smallest minimal defining quartet set), which for a binary tree is just the number of internal edges of $T$. Furthermore, \citet[Theorem~6.8.8]{semple2003pol} provide a quick method for constructing minimal defining sets of quartets that define binary phylogenetic trees. 

Let $V\subset U$ be such that $V=\{i,j,k,l\}$, where these elements are distinct. Consider three random variables $Q_{ik,jl}$, $Q_{il,jk}$ and $Q_{ij,kl}$ as defined in (\ref{eq:Qij}). By Theorem~\ref{thm:ineq}, if a tree model holds, then the mean of one of the three will be zero and the other two means will be equal up to sign. So  these $Q_{I,J}$ can be used to test the algebraic constraints in Proposition~\ref{prop:fixedpot}.

Here we focus on testing the vanishing tetrads, i.e. testing whether the quartet $ q_{ij,kl}$ is displayed in $T$ given the data. To test a particular binary tree $T$, a set $\mathcal{Q}_{T}$ is required, i.e.\ a set of quartets $\mathcal{Q}$ that determines $T$. The number of edges of $T$ is $2m-3$ and so the Gaussian latent tree model on $T$ has codimension ${m\choose 2}-(2m-3)$. This means that to test a model, we need to work with quartet systems $\mathcal{Q}_{T}$ of size quadratic in $m$. On the other hand if we believe that the data come from a latent tree model, then to only find the corresponding tree $T$ we can work with any minimal quartet system determining $T$, and these are of size $m-3$ (the number of internal edges). This makes a big difference for larger trees. 

 In practice, one may wish to select $\mathcal{Q}_{T}$ such that it is minimal (of size ${m\choose 2}-(2m-3)$), i.e.\ it contains no redundant  quartets, because otherwise the covariance of minors matrix may be close to being singular; see \cite{bollen1993}. However, there may not always be an obvious reason for selecting one minimal defining quartet set $\mathcal{Q}_{T}$  over another. In such cases one approach is to randomly select a number of sets to assess the robustness of the results; see \cite{bollen1993}. For each $q_{ij,kl} \in \mathcal{Q}_{T}$ consider the corresponding $Q_{ij,kl}$ as in (\ref{eq:Qij}) and define $ Q_{T}=[Q_{ij,kl}]$ to be the vector of these $Q_{ij,kl}$. We write $\hat Q_{ij,kl}$ for the sample means of the observations of $\hat Q_{ij,kl}$. Since $\hat Q_T$ is a consistent estimator of $Q_T$ (see \citet{pentads2007}), as the sample size $n$ tends to infinity any tree $T$ is uniquely identified by the $i,j,k,l$ such that $E(\hat{Q}_{ij,kl})=0$. 

To standardize the data we use the sample covariance matrix  $\hat{\Sigma}_{Q_{T}}$ which has dimension $p = \vert Q_{T}\vert$. Alternatively we can use its proxy $\tilde{\Sigma}_{Q_{T}}$. This can then be obtained by recycling ${\rm cov} (W^{(2)})$ computed in Section~\ref{sec:moments} and using (\ref{eq:covwhite}) substituting $C$ for the sample covariance of original variables $\hat\Sigma$. The matrix $\tilde{\Sigma}_{Q_{T}}$ can be obtained much more efficiently than $\hat{\Sigma}_{Q_{T}}$. An appropriate simultaneous test statistic (\ref{eq:teststat}) is provided in \citet{bollen1993}, 
\begin{equation}\label{eq:teststat}
\mathcal{T} \;\;=\;\; \hat{Q}^t_T\,\hat{\Sigma}^{-1}_{Q_T}\,\hat{Q}_T,
\end{equation}
where $A^t$ is the transpose of $A$. Given $\mathcal T$ is constructed with $p$ algebraically independent quartets, the asymptotic distribution of this test statistic is $\chi^2$-distribution with $p$ degrees of freedom. Compare (\ref{eq:teststat}) with \citet[(20)]{bollen1993} where their $\Sigma_{tt}$ is the covariance of $\sqrt{n}\hat{Q}_T$. Here the sample size $n$ is incorporated implicitly through $\hat{\Sigma}^{-1}_{Q_T}$ so (\ref{eq:teststat}) provides a significance test for the equality constraints in (\ref{eqn:tetrad}), where the required moments of $Q_{I,J}$ are given in Section~\ref{sec:moments}. This provides a quick method for assessing whether a Gaussian data set appears consistent with the algebraic constraints associated with any tree model.

In deriving the asymptotic distribution in (\ref{eq:teststat}) we implicitly assume that the true covariance matrix is a sufficiently regular point of the given tree model. In practice, it is enough to assume that the true covariance matrix contains no zeros; see Section 5 in \cite{drton2013wald} and \cite{DLWZtreesSBIC}. 

Hypothesis testing for vanishing tetrads can be used for both confirmatory tetrad analysis and for exploratory tetrad analysis. There are many algorithms for obtaining candidate trees, for instance see \citet{junker2011analysis,sung2009algorithms} for  surveys of methods. However, often there is no way to assess the suitability of the finally chosen tree. Confirmatory tetrad analysis takes a candidate tree and provides an absolute rather than relative value as to how well the data supports the purported tree. 

In the case of a large tree it is infeasible to test all quartets at once. On the other hand, it is straightforward and very stable to test single quartets or a small subset of them. One advantage of this approach is that it allows us to identify easily certain macrostructures of the tree which may lead to more robust techniques for finding the underlying tree. We now illustrate confirmatory and exploratory techniques both for simulated data and some linguistics data sets.

\subsection{Basic simulations for the method of quartets}\label{sec:basiccomp}

In this section we provide a basic analysis of the methods discussed in the previous section. The only difference from the previous applications of this method in other contexts is that in (\ref{eq:teststat}) we explicitly replaced the sample covariance of the tetrads with $\tilde{\Sigma}_{Q_{T}}$ as explained in Section~\ref{sec:quarttest}. The data in our simulations come from the quintet tree model $12|5|34$; c.f. Figure~\ref{fig:quintetlang}. 

We first randomly choose the true covariance matrix $C$ by sampling the edge correlations uniformly from the interval $[1/2,1]$. Given this random true covariance matrix, we can now repeat the following evaluation procedure 10{,}000 times. We sample $n=60$ ($5$ times the dimension of the model) points from the given distribution $C$. In this scenario, standard packages that might be used to find the maximum likelihood estimate, such as the \texttt{sem} package \citep{sem_package}, are unstable. On the other hand any set of quartets can be easily tested, and this does not even require any fitting of the model. Moreover, the sample distribution of the test statistic is already very close to the asymptotic distribution. Quite surprisingly this proximity remains true even when the sample size is only about twice the dimension of the model. Of course, in this case, the power of the test will be much lower.

In Figure~\ref{fig:results} we show the simulated values of test statistics of the form (\ref{eq:teststat}) compared with their theoretical asymptotic distributions. Figure~\ref{fig:results}(a) depicts the statistic built on a single tetrad constraint for the quartet $12|34$. This constraint holds for the data generating distribution and therefore the test statistic is expected to have asymptotic $\chi^2$-distribution with one degree of freedom. We see that the histogram is very close to the theoretical distribution. For comparison, in  Figure~\ref{fig:results}(b) we show the sample distribution of the same test statistic for the quartet $13|24$. This constraint does not hold for the data generating distribution and we see that the sample distribution of the corresponding test statistic is very far from $\chi^2_1$. The test statistic can be easily set up for any subset of quartets. In Figure~\ref{fig:results}(c) we plot the test statistic to test two quartets $12|35$ and $15|34$. This is the minimal set of quartets that identifies the quintet tree $12|5|34$. This means that these two particular quartet will not be simultaneously satisfied for any other tree model  . Again, the sample distribution lies very close to the asymptotic distribution, which in this case is $\chi^2_2$.

In Figure~\ref{fig:results}(d) we test simultaneously a minimal set of quartets defining the quintet tree model; these are: $12|34$, $12|35$, $15|34$. In this case the sample distribution of the test statistic also lies close to $\chi^2_3$ with a slightly smaller variance. The reason for that is that the true distribution is closer to a mixture of $\chi^2$-distributions. As a result, the test based on our statistic is typically more conservative. To obtain a better understanding of its performance we compare it with the structural expectation-maximization algorithm \cite{friedman2002structural} as applied to Gaussian latent tree models. This algorithm tries to find the tree that gives the maximum value of the likelihood function. However, like the standard expectation-maximization algorithm, it often gets stuck in a local maximum. In our simulations we generated $100$ data sets from the given quintet model. If the sample size $n=60$, then for our particular choice of a correlation matrix with all edge correlations equal to $0{.}7$, we obtained the correct tree only $68$ out of $100$ times. On the other hand, our tetrad method always confirms the correct tree on any significance level smaller than $0{.}1$. If $n=200$ then the structural expectation-maximization algorithm was correct $99$ out of $100$ times, and again our quartet method was always correct. We emphasize that in a less ideal situation, for example, when some edge correlations are small, or in the presence of some partial misspecification, the structural expectation-maximization algorithm will perform poorly because the likelihood function is less stable. In contrast, our computations show that the quartet method tends to be much more robust.

\subsection{Exploratory tetrad analysis example:\ linguistics}
\label{sec:eta}
Consider now the linguistic data set from \citet{shi14gtc}. This comprises phonetic functional spectrogram data from five Romance languages: French, Italian, Portuguese, and two forms of Spanish, namely American and Iberian. Acoustic data have provided new insights into language developments (e.g.\ \citet{bou13ara}, \citet{ast10lpa}). Here the evolutionary dependencies between spoken numbers is studied with each extant language treated as a leaf vertex. The high dimensional spectrogram data is projected from 8100 dimensions to 15 dimensions using a variation of canonical variate analysis, the full details of which can be found in \citet{shi14gtc}. Each of the 15 canonical components projects the mean word data to obtain 15 new data sets referred to as canonical scores. Each canonical component accounts for a particular combination of phonetic variation and each set of canonical scores is considered independently. This gives us the flexibility to hypothesize different evolutionary relationships for different aspects of the speech. For each set of canonical scores a $5\times 5$ covariance matrix is calculated between the five languages. Royston's multivariate normality test \citep{roy83sta} does not reject Gaussianity at the $0.01$ level for any of these 15 sets of scores. 

We sampled $10^{5}$ covariance matrices from the inverse-Wishart posterior for each of the sample covariances $\hat{\Sigma}_{1}, \ldots, \hat{\Sigma}_{15}$. We then performed a tree-compatibility test with respect to the positivity constraint implied by the triangle inequalities in Theorem \ref{thm:ineq}(2) for each canonical component. We identify four such components, the first, fourth, sixth, and second, with high posterior probabilities, respectively 1, 0.89, 0.77 and 0.74, which warrant further investigation.

Considering the quintet tree in Fig.~\ref{fig:quintetlang}, there are 15 different labelled binary trees to test. In order to test a particular configuration of labels we construct a set of minimal defining quartets $\mathcal{Q}$ for the quintet tree as referenced in Section~\ref{sec:quarttest}; in the case of the quintet tree this smallest minimal set is two. 

For each of the four dimensions of interest, using the sampling distributions given in Section~\ref{sec:moments} and the test statistic (\ref{eq:teststat}) with two degrees of freedom, a p-value can be calculated for each of the 15 non-isomorphic trees with languages as leaves. To retain an overall significance rate of less than $0.05$ a Bonferroni correction \citep{dunn1961multiple} is applied such that the significance level is set at $0.05/15 \approx 0.0033$ per test. If more than one tree is not rejected then the candidate tree proposed by exploratory tetrad analysis is that with the highest p-value. We find that multiple trees exceed the threshold for all four components. The highest  p-values for the first, second, fourth and sixth components were 0.524, 0.960, 0.775 and 0.902 respectively relating to the candidate trees: $12\vert 4\vert 35$, $13\vert 5\vert 24$ ,$14\vert 2\vert 35$, and $23\vert 1\vert 45$ respectively with coding 1 = French, 2 = Italian, 3 = Portuguese, 4 = American Spanish, 5 = Iberian Spanish. 

For illustration we focus on the candidate tree for the second component, which is displayed in Fig~\ref{fig:quintetlang}. It is known from the analysis and expert interpretation in \citet{shi14gtc} that this component is likely to relate to variation in vowel sounds, nasality, and the lip rounding of language speakers. By isolating these phonetic features and identifying an accompany tree that fits the data we can gain insights which may have otherwise been obscured. For example, from this particular analysis we could hypothesize that the differences in nasality of Italian and French evolved independently conditional on the common ancestor of Spanish and Portuguese. In combination with expert judgement, such statements can provide good starting points for further analyses of these features in relation to a specified tree.

\subsection{Confirmatory tetrad analysis example:\ biology}
\label{sec:cta}

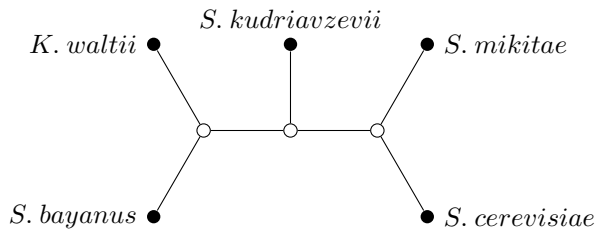
\begin{figure}[ht!]
\centering
\tikzstyle{vertex}=[circle,fill=black,minimum size=5pt,inner sep=0pt]
\tikzstyle{hidden}=[circle,draw,minimum size=5pt,inner sep=0pt]
\begin{tikzpicture}[scale=.9]
  \node[hidden] (h2) []{};  
  \node[hidden] (h1) at (-1.27,0) []{};
  \node[vertex] (x5) at (0,1.27) [label=above:$S.\ kudriavzevii$]{};
  \node[hidden] (h3) at (1.27,0) []{};
  \node[vertex] (x1) at (-2,1.27) [label=left:$K.\ waltii $]{};
  \node[vertex] (x2) at (-2,-1.27) [label=left:$S.\ bayanus$]{};
  \node[vertex] (x3) at (2,1.27) [label=right:$S.\ mikitae$]{};
  \node[vertex] (x4) at (2,-1.27) [label=right:$S.\ cerevisiae$]{};
  \path (h1) edge (h2)
		(h2) edge (h3)
		(h1) edge (x1)
 		(h1) edge (x2)
  		(h3) edge (x3)
 		(h3) edge (x4)
  		(h2) edge (x5);
\end{tikzpicture}
\caption{Quintet tree $T_{5}$ of yeast species \citep{marcet2009tree}.} \label{fig:yeastquint}
\end{figure}

We next consider a data set consisting of growth curves for five yeast species each observed in the same 96 environments, each species with at least two replicates. The growth was recorded at approximately six minute intervals over a period of just over 26 hours. These species have been studied before \citep{marcet2009tree} and a phylogeny has been suggested as in Fig.~\ref{fig:yeastquint}. However, \citet{lib11mdi} hypothesize that yeast species S.\ bayanus is a hybrid involving S.\ cerevisiae. This alternative hypothesis would violate the tree assumption. Previous research has indicated that for studied yeast species there is positive correlation between growth-related phenotypic variation and genotypic phylogenetic relationships, e.g.\ \cite{lit09pgd,war11tvy}. Thus, this leads us to consider the yeast growth-curve data to investigate evolutionary questions. We carried out a confirmatory tetrad analysis to assess whether the proposed tree structure in \citet{marcet2009tree} was reflected in any aspects of the growth data.

To pre-process the data, a smoothed cubic spline basis was fitted to each growth vector resulting in a set of functional data objects which were then regularly evaluated to obtain comparable discretized representations. Mean vectors were then calculated for each species and environment and then these were standardized to remove mean environmental effects. We then performed a principal component analysis across species to identify the core variability of the growth curves. Note that the first four components account for over 99\% of variability. More detailed analysis, not reported here, can help interpret these components. For example, the first component relates only to growth variation in hours 10 to 26, whereas the second component relates to growth variation peaking at 12 hours with opposite growth variation from 18 hours onwards. 

For each of the mean species projections in these dimensions, the sample covariance matrix was constructed. As a first step, the inverse-Wishart approach specified in (\ref{eq:indic}) was implemented. Recall that the tripod constraints are tree-compatibility constraints and thus require no tailoring to a specific $T$. Hence, these can be utilized very simply to narrow the list of components to test as part of a confirmatory tetrad analysis. The tree-compatibility for the first four components were 31\%, 2\%, 18\%, and 3\% respectively. Thus, we consider the first and third components worth investigating further via confirmatory tetrad analysis for $T$-compatibility.

The results of the confirmatory tetrad analysis for $T_{5}$-compatibility (see Fig.~\ref{fig:yeastquint}) gave p-values of 0.721 and 0.955 for the first and third components respectively. To double check these results we repeated the test using the bootstrapping strategy outlined in \citet{bol92bgf}. The results were very similar with p-values of 0.729 and 0.921 respectively. The confirmatory tetrad analysis and inverse-Wishart simulation results both gave upper bounds on $T_{5}$-compatibility, but on balance we concluded that the first and third components were $T_{5}$-compatible. Therefore, the class of Gaussian latent tree models did appear suitable for modelling some aspects of these yeast species' growth curves. However, for features relating to components 2 and 4, there is some evidence to support the exploration of a wider model class that could accommodate the hybrid hypothesis described in \citet{lib11mdi}.

\section{Discussion}

Understanding the complete description of the correlation space associated with Gaussian latent tree models opens up a number of useful tools for assessing tree-compatibility either on a class basis or for a specified tree. Some of the methods described in this paper are particularly useful as part of an exploratory analysis for defining the relevant model search space, whereas others are ideal as a final step to check the conclusions of a model search. The complete semialgebraic structure of the correlation space has not been utilized elsewhere for assessing tree-compatibility of data, though the positivity constraint has been used previously, see \citet{shi14gtc}. Incorporating a prior such as the inverse-Wishart and sampling from the posterior distribution allows for probabilistic conclusions about the model. This provides a more nuanced answer than a simple assessment of inequalities via the plugging in of covariance point estimates, and enables two or more incompatible but plausible trees to be compared. 

One important practical consideration is the scalability of these methods. Techniques employing the semialgebraic constraints can be adapted to larger number of variables reasonably well. For a confirmatory tetrad analysis the biggest computational cost is the calculation of the covariance of minors, which for $p$ observed random variables has dimension of order $p^4$ which can become prohibitive. For example, if $8$GB of RAM is allocated for a single matrix, the limit of $p$ is approximately $25$ even if redundant rows and columns are removed from the matrix. However, much larger $p$ can be considered by calculating the relevant statistics for each quartet marginally. Then the covariance matrix of minors has dimension of only $36$ and the memory can be released once each quartet has been tested. In either case, the final memory requirement could further be reduced with smart programming taking advantage of symmetries and sparseness. In a similar vein, to extend the scope of exploratory tetrad analysis to a greater number of variables, one strategy is to only assess single quartets in the first instance and use these results to reduce the set of possible trees worth considering. Given the effectiveness of the quartet testing as demonstrated with even small sample sizes, this approach seems sensible and to have significant advantage over methods that require a whole model to be tested at once.

\section*{Acknowledgement}
We are grateful to both referees for critical comments that substantially improved the presentation of the paper. Nathaniel Shiers acknowledges the support of the ESRC. 
Piotr Zwiernik was supported by the European Union 7th Framework Programme. 
The authors wish to thank Pantelis Hadjipantelis for preprocessing the linguistic data, John S. Coleman for interpretation of the linguistic analysis, and Chris Knight for provision of the yeast data.

\appendix

\section{Proofs}

\begin{proof}[of Theorem \ref{thm:ineq}]Assume first that all correlations $\rho_{ij}$ are strictly positive, that is $R\in {\rm PO}_{+}(V)$ or $R\in {\rm PO}_{+}(T)$. We use the fact that ${\rm PO}_{+}(V)$ is  isomorphic to the space of tree metrics, whose constraints are given in Theorem~\ref{th:buneman} and Corollary~\ref{cor:buneman}. Translating these constraints via $d_{ij}=-\log(\rho_{ij})$ gives exactly the constraints  in the proposed theorem. These constraints describe a closed set, which is the smallest closed set containing ${\rm PO}_{+}(V)$. So it is enough to show that the closure of ${\rm PO}_{+}(T)$ is equal to ${\rm PO}(T)$. This follows from the fact that ${\rm PO}(T)$ is a toric cube and, by \citet[Theorem~1]{toriccubes}, every toric cube is equal to the closure of its interior. 
\end{proof}

\begin{proof}[of Theorem \ref{th:ineqpos}]
If $R\in M(T)$ then each $\rho_{ij}$ has representation (\ref{eq:rhoijpath}). Thus $|\rho_{ij}|=\prod_{e\in {\rm ph}(ij)}|\rho_{e}|$ and hence $R'$ also lies in ${\rm PO}(T)$. To show that $\rho_{ij}\rho_{ik}\rho_{jk}\geq  0$ consider  the induced subtree over $i,j,k$, that is, the smallest connected subgraph of $T$ containing vertices $i,j,k$. This subtree necessarily has a unique vertex $v$ that lies on the intersection of paths ${\rm ph}(ij)$, ${\rm ph}(ik)$ and ${\rm ph}(jk)$. Moreover, by (\ref{eq:rhoijpath}),
$$
\rho_{ij}\rho_{ik}\rho_{jk}=\prod_{e\in {\rm ph}(ij) }\rho_{e}\prod_{e\in {\rm ph}(ik) }\rho_{e}\prod_{e\in {\rm ph}(jk) }\rho_{e}=\prod_{e\in {\rm ph}(iv) }\rho_{e}^{2}\prod_{e\in {\rm ph}(jv) }\rho_{e}^{2}\prod_{e\in {\rm ph}(kv) }\rho_{e}^{2}\geq  0.
$$
To prove the reverse implication, we note that every correlation matrix in ${\rm PO}(T)$ has (after permuting rows and columns) a block diagonal structure with strictly positive elements in each block. Consider first the case when all elements of $R$ are non-zero, that is,  $R'$ has strictly positive entries. Distinguish one node in $V$ and label it as $1$. Let $D$ be a diagonal matrix such that $D_{ii}=-1$ if $\rho_{1i}<0$ and $D_{ii}=1$ if $\rho_{1i}>0$. If $R\in M(T)$ then also $DRD$ lies in $M(T)$ because $M(T)$ is invariant with respect to all diagonal transformations. Moreover, it holds that $R'=DRD$ because $D_{11}D_{ii}\rho_{1i}=|\rho_{1i}|$ for all $i\in V\setminus \{1\}$ and $D_{ii}D_{jj}\rho_{ij}=|\rho_{ij}|$ for $i,j\in V\setminus \{1\}$. This last equality follows from our assumption that $\rho_{1i}\rho_{1j}\rho_{ij}\geq  0$ so that the sign of $\rho_{1i}\rho_{1j}$ is equal to the sign of $\rho_{ij}$. Now, since $R'\in {\rm PO}(T)\subset M(T)$ and $R=DR'D$ we also have that $R\in M(T)$. The analysis of the case when $R$ is block diagonal will be omitted.
\end{proof}

\bibliographystyle{biometrika}
\bibliography{tree_models}
\newpage
\begin{figure}[t!]
\centering
\subfigure[single tetrad]{\includegraphics[scale=0.55]{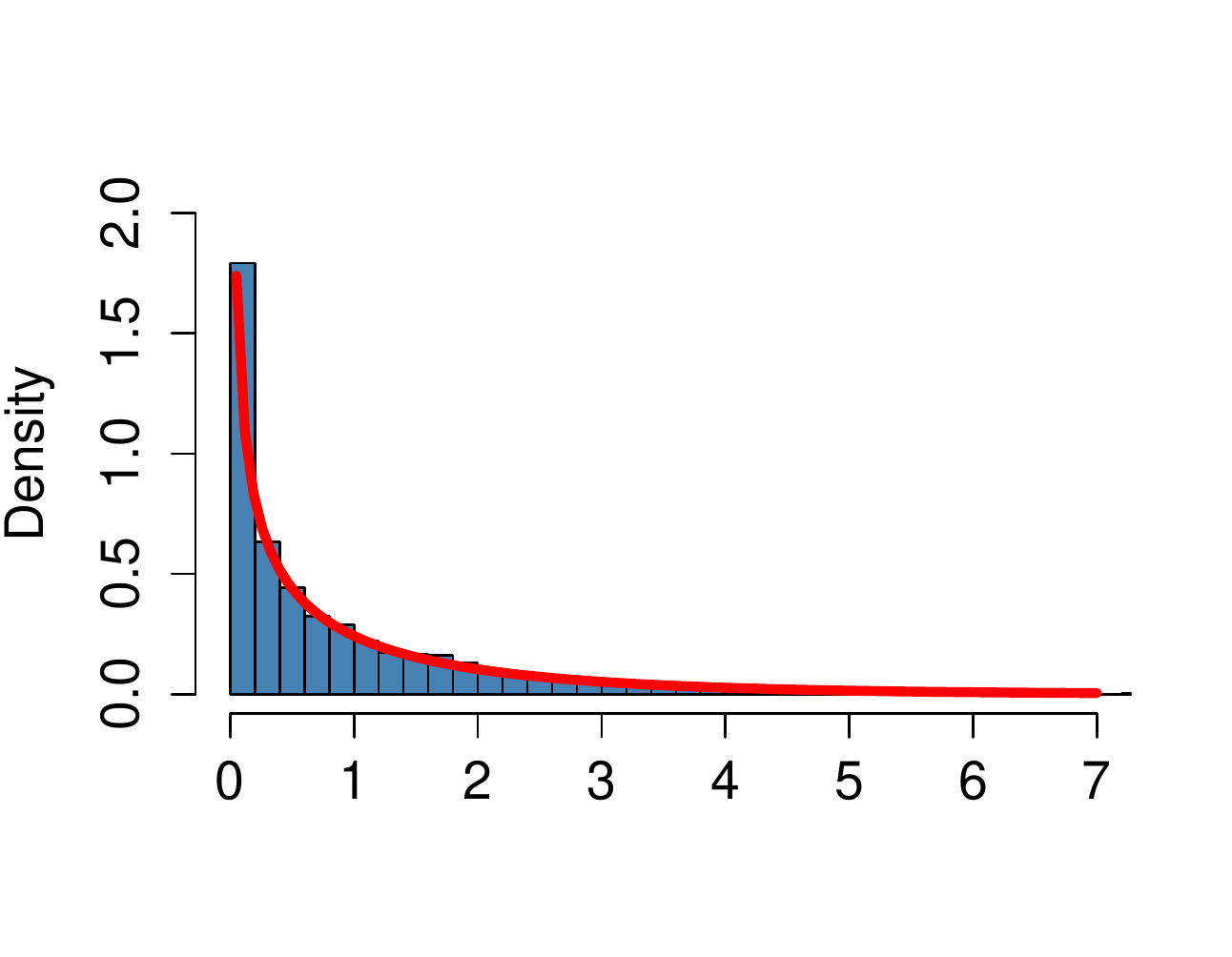}}
\subfigure[single false tetrad]{\includegraphics[scale=0.55]{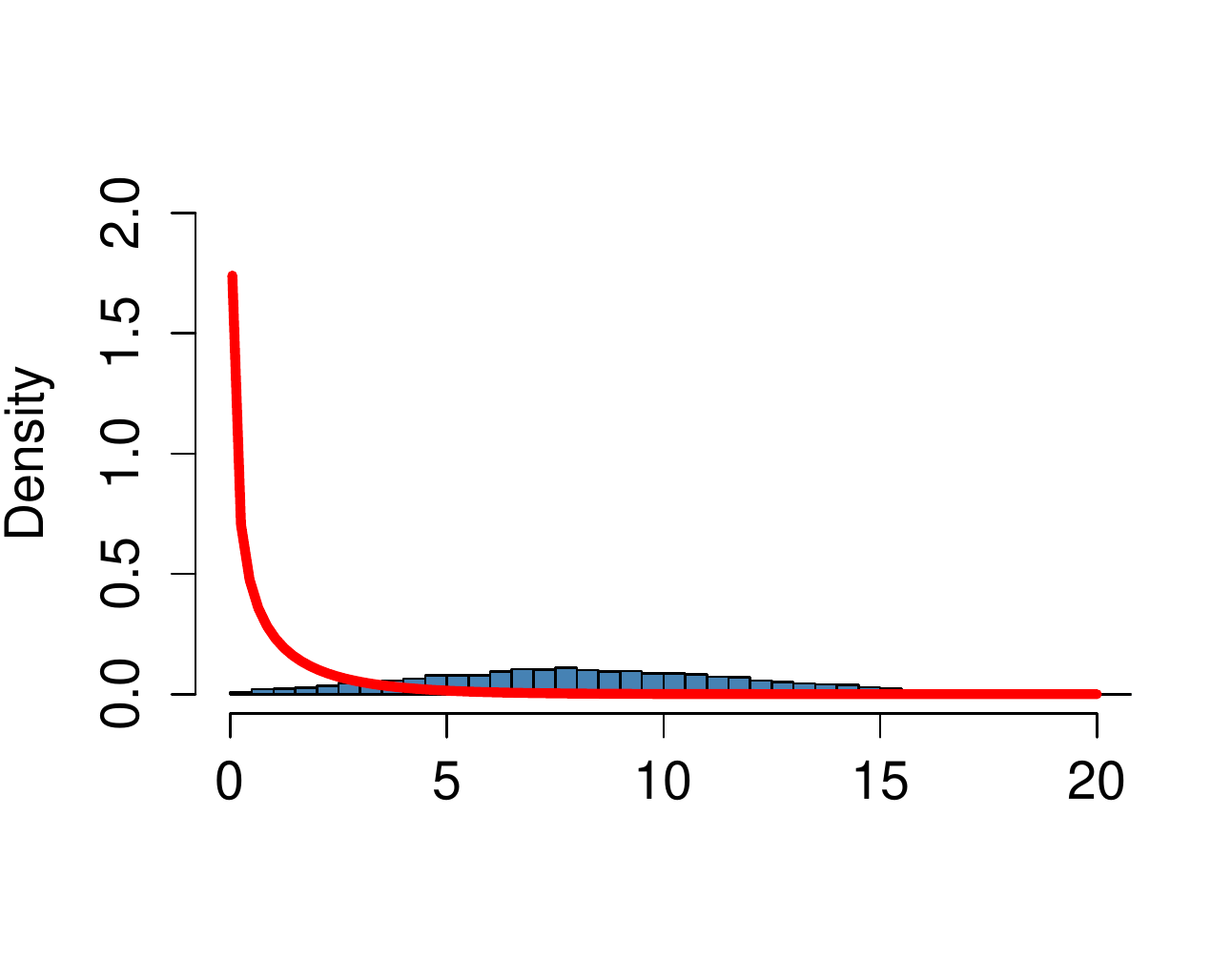}}\\
\subfigure[two tetrads]{\includegraphics[scale=0.55]{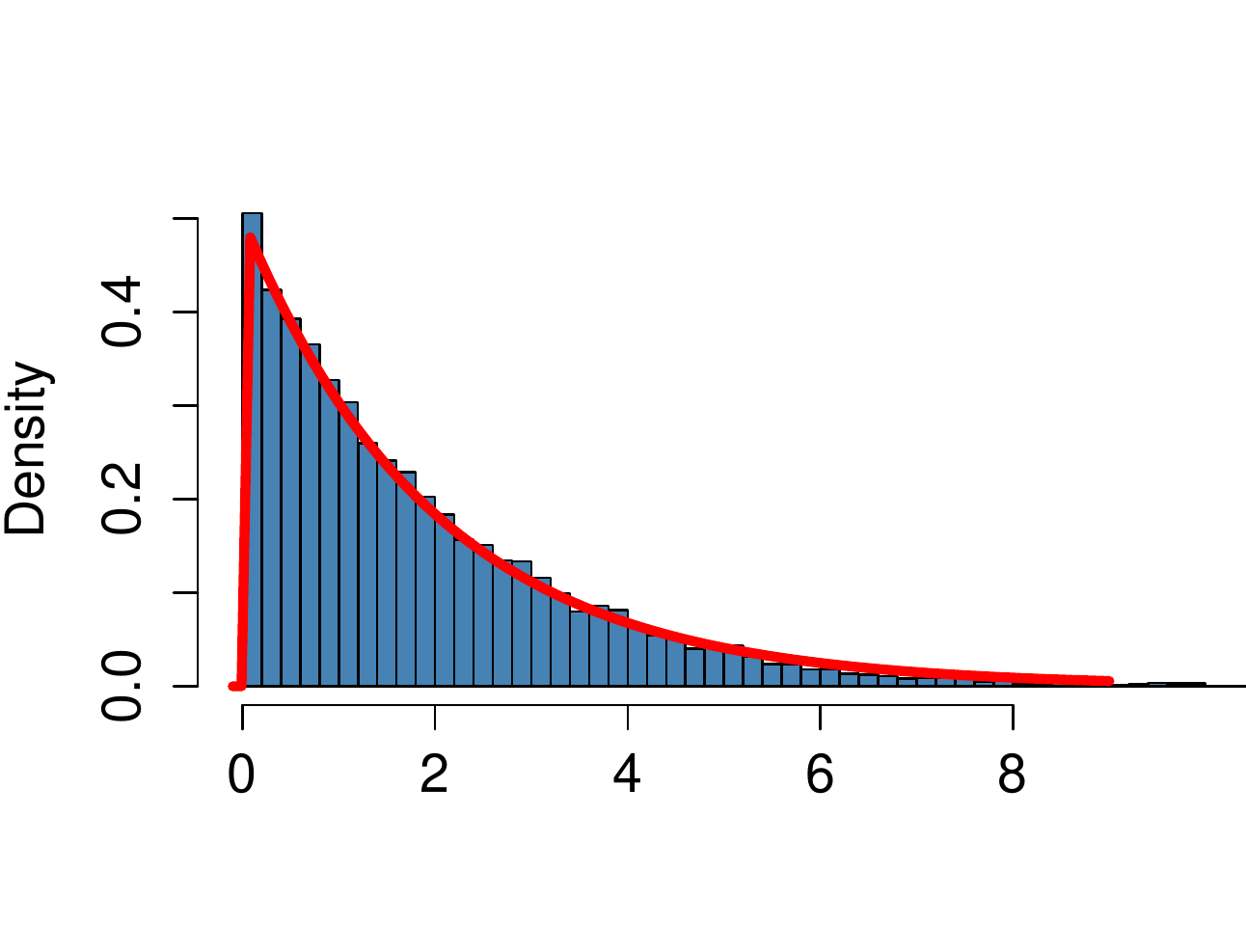}} 
\subfigure[defining tetrads]{\includegraphics[scale=0.55]{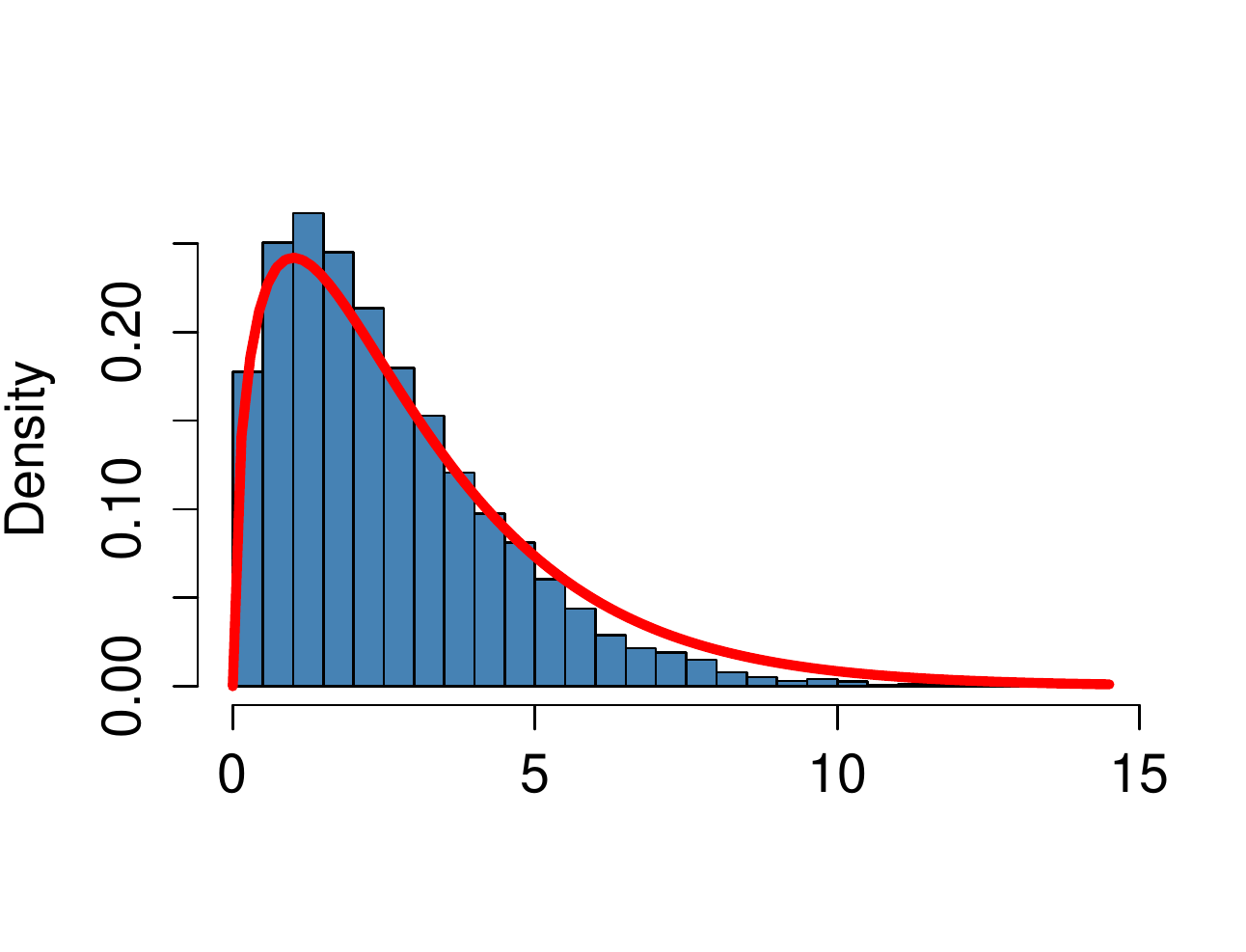}} 
\caption{Illustration of the simulations in Section~\ref{sec:basiccomp}. Sixty observations are generated from a random matrix in the tree model for the quintet tree $12|5|34$ and the corresponding test statistic is computed. This procedure is iterated $10{,}000$ times. In each figure we compare the sample distribution of a test statistics against its theoretical distribution. In (a) we test a single tetrad $12|34$. In (b) we test a single false tetrad $13|24$. In (c) we test  two tetrads $12|35$, $15|34$, and in (d) we test a minimal set of quartets defining the true quintet tree. The solid lines are densities of $\chi^2_1$, $\chi^2_1$, $\chi^2_2$, and $\chi^2_3$ respectively.}
\label{fig:results}
\end{figure}

\end{document}